\documentstyle[aps,prl,epsf]{revtex}
\begin{document}
\twocolumn[\hsize\textwidth\columnwidth
\hsize\csname@twocolumnfalse\endcsname
\draft
\title{Observation of the distribution of molecular spin states\\ by
resonant quantum tunneling of the magnetization}
\author{W. Wernsdorfer$^\ast$, T. Ohm$^\dag$, C.~Sangregorio$^\ddag$, R.~
Sessoli$^\ddag$, D.~Mailly$^\S$, and C.~Paulsen$^\dag$}
\address{
$^\ast$Laboratoire Louis N\'eel, CNRS, BP 166, 38042 Grenoble Cedex
9, France\\
$^\dag$CRTBT, associ\'e \`a l'Universit\'e Joseph Fourier, CNRS, BP 166,
38042 Grenoble Cedex 9, France\\
$^\ddag$Department of Chemistry, University of Firenze, via Maragliano 72,
50144
Firenze, Italy\\
$^\S$LMM, CNRS, 196 Avenue H. Ravera, 92220 Bagneux, France
}

\date{}

\maketitle

\begin{abstract}
Below 360~mK, Fe$_8$ magnetic molecular clusters are in the pure
quantum relaxation regime and we show that the predicted
``square-root time'' relaxation is obeyed, allowing us to develop a
new method for watching the evolution of the distribution of molecular
spin states in the sample. We measure as a function of applied field
$H$ the statistical distribution $P(\xi_H)$ of magnetic energy bias
$\xi_H$ acting on the molecules. Tunneling initially causes rapid
transitions of molecules, thereby ``digging a hole'' in $P(\xi_H)$
(around the resonant condition $\xi_H$~= 0). For
small initial magnetization values, the hole width shows an intrinsic
broadening which may be due to nuclear spins.
\end{abstract}
\bigskip

\vskip1pc]

\narrowtext

Strong evidence now exists for thermally-activated quantum tunneling
of the magnetization (QTM) in magnetic molecules such as Mn$_{12}$ac
and Fe$_8$ \cite{Novak95,Paulsen95,Friedman96,Thomas96,Sangregorio97}.
Crystals of these materials can be thought of as ensembles of
identical, iso-oriented nanomagnets of net spin $S$~= 10 for both
Mn$_{12}$ac and Fe$_8$, and with a strong Ising-like anisotropy. The
energy barrier between the two lowest lying spin states with $S_z$~=
$\pm$10 is about 60~K for Mn$_{12}$ac and 25~K for Fe$_8$
\cite{Sessoli93,Barra96}. Theoretical discussion of
thermally-activated QTM assumes that thermal processes (principally
phonons) promote the molecules up to high levels, not far below the
top of the energy barrier, and the molecules then tunnel inelastically
to the other side. The transitions are therefore almost entirely
accomplished via thermal excitations.

At temperatures below 360~mK, Fe$_8$ molecular clusters display a
clear crossover from thermally activated relaxation to a temperature
independent quantum regime, with a pronounced resonance structure of
the relaxation time as a function of the external field
\cite{Sangregorio97}. This can be seen for example by hysteresis loop
measurements (Fig.~\ref{fig1}). In this regime only the two lowest
levels of each molecule are occupied, and only ``pure'' quantum
tunneling through the anisotropy barrier can cause direct transitions
between these two states. It was surprising however that the observed
relaxation of the magnetization in the quantum regime was found to be
non-exponential and the resonance width orders of magnitude too large
\cite{Sangregorio97,Ohm98}. The key to understanding this seemingly
anomalous behavior now appears to involve the ubiquitous hyperfine
fields as well as the (inevitable) evolving distribution of the weak
dipole fields of the nanomagnets themselves \cite{Prokofev98}.

In this letter, we focus on the low temperature and low field limits,
where phonon-mediated relaxation is astronomically long and can be
neglected. In this limit, the $S_z$~= $\pm$10 spin states are coupled by a
tunneling
matrix element $\Delta_{\rm tunnel}$ which is estimated to be about
10$^{-8}$~K \cite{Prokofev98}. In order to tunnel between these
states, the magnetic energy bias $\xi_H=g\mu_BSH$ due to the local
magnetic field $H$ on a molecule must be smaller than $\Delta_{\rm 
tunnel}$ implying a local field smaller than $10^{-9}$~T for Fe$_8$
clusters. Since the typical intermolecular dipole fields are of the
order of 0.05~T, it seems at first that almost all molecules should be
blocked from tunneling by a very large energy bias. Prokofev and Stamp
have proposed a solution to this dilemna by assuming that fast dynamic
nuclear fluctuations broaden the resonance, and the gradual adjustment
of the dipole fields in the sample caused by the tunneling, brings
other molecules into resonance and allows continuous relaxation
\cite{Prokofev98}. A crucial prediction of the theory is that at a
given longitudinal applied field $H$, the magnetization should relax
at short times with a square-root time dependence:
\begin{equation}
	M(H,t)=M_{\rm in}+(M_{\rm eq}(H)-M_{\rm in})\sqrt{\Gamma_{\rm
	sqrt}(\xi_H)t} \label{eq1}
\end{equation}
Here $M_{\rm in}$ is the initial magnetization at time $t$~= 0 (i.e. after a
rapid field change), and $M_{\rm eq}(H)$ is the equilibrium
magnetization. The rate function $\Gamma_{\rm sqrt}(\xi_H)$ is
proportional to the normalized distribution $P(\xi_H)$ of energy bias
in the sample:
\begin{equation}
	\Gamma_{\rm sqrt}(\xi_H)=c\frac{\Delta^2_{\rm tunnel}}{\hbar}P(\xi_H)
	\label{eq2}
\end{equation}
where $\hbar$ is Planck's constant and $c$ is a constant of the order
of unity which depends on the sample shape. If these simple relations
are true, then measurements of the short time relaxation as a
function of the applied field $H$ gives experimentalist a powerful new
method to directly observe the distribution $P(\xi_H)$. Indeed the
predicted $\sqrt{t}$ relaxation (Eq.~(\ref{eq1})) has been seen in
preliminary experiments on fully saturated Fe$_8$ crystals \cite{Ohm98}.
We show here that it is accurately obeyed for saturated and
non-saturated samples (Fig.~\ref{fig2}) and we find that a remarkable
structure emerges in $P(\xi_H)$ as presented in the following.

In order to carefully study $P(\xi_H)$ and its evolution as the sample
relaxes, we have developed a unique magnetometer consisting of an
array of micro-SQUIDs \cite{Wernsdorfer97,Wernsdorfer96} on which we
placed a single crystal of Fe$_8$ molecular clusters. The SQUIDs
measure the magnetic field induced by the magnetization of the
crystal (see inset of Fig.~\ref{fig1}). The advantage of this
magnetometer lies mainly in its high sensitivity and fast response,
allowing short-time measurements down to 1~ms. Furthermore the
magnetic field can be changed rapidly and along any direction.

Figure~\ref{fig2} shows a typical set of relaxation curves plotted
against the square-root of time. However instead of saturating the
sample before each relaxation measurement so that the initial
magnetization $M_{\rm in}=M_s$ as described in \cite{Ohm98}, these
measurements were made by rapidly quenching the sample from 2~K in
zero field (ZFC), i.e. for an initial magnetization $M_{\rm in}$~= 0.
The quench takes approximately one second and thus the sample does
not have time to relax, either by thermal activation or by quantum
transitions, so that the high temperature ``thermal equilibrium''
spin distribution is effectively frozen in. Once the temperature is
stable (in this case 40~mK) a measuring field is applied, the timer
is set to $t$~= 0, and the relaxation of the magnetization is recorded
as a function of time. The entire procedure was repeated for {\em
each} measuring field shown in Fig.~\ref{fig2}. As can be seen for
short times $t<100$~s the square root relaxation is well obeyed. Note
that all curves extrapolate back to $M$~= 0 at $t$~= 0. A fit of the
data to Eq.~(\ref{eq1}) determines $\Gamma_{\rm sqrt}$.

A plot of $\Gamma_{\rm sqrt}$ vs. $H$ is shown in Fig.~\ref{fig3} for
the zero field cooled data, as well as distributions for three other
values of the initial magnetization which were obtained by quenching
in small fixed fields (field cooled FC magnetization). The distribution
or an initial magnetization close to the saturation value is clearly
the most narrow reflecting the high degree of order starting from this
state. The distributions become more broad as the initial
magnetization becomes smaller reflecting the random fraction of
reversed spins. The small satellite bumps are due to flipped nearest
neighbor spins on the tri-clinic lattice as seen in computer
simulations \cite{Ohm98,Prokofev98JLTP}.

We can exploit this technique of measuring $P(\xi_H)$ in order to
observe the evolution of molecular states in the sample during
relaxation by quantum tunneling of the magnetization. We first field
cooled the sample (thermally anneal) as described above in order to
obtain the desired initial magnetization state. Then after applying a
field $H_t$, we let the sample relax for a time $t_t$, which we call
``tunneling field'' and ``tunneling time'' respectively. During the
tunneling time, a small fraction of the molecular spins tunnel and
reverse their direction. Finally, we applied a small measuring field
and record the short time relaxation which again can be fit to a
square root law (Eq.~(\ref{eq1})) yielding $\Gamma_{\rm sqrt}$. The
entire procedure is repeated many times for other measuring fields in
order to probe the distribution as a function of field $H$, and thus
we obtain the distribution $P(\xi_H,H_t,t_t)$ which we call a
``tunneling distribution''.

Figure~\ref{fig4} shows tunneling distributions for field $H_t$~= 0
and for tunneling times between 1 and 250~s for the case that the
initial magnetization starts from the
fully saturated state. Note the rapid depletion of molecular spin
states around the resonant field $H_t$~= 0 and how quickly the
depletion depth and width increase with tunneling time. In effect, a
``hole is dug'' into the distribution function around $H_t$. The hole
arises because only spins in resonance can tunnel. The hole is spread
out because as the sample relaxes, the internal fields in the sample
change such that spins which were close to the resonance condition may
actually be brought into resonance. Notice however that ``wings'' are
created on each side of the hole because other spins are pushed
further away from resonance. These features are in good agreement with
Monte Carlo simulations of the relaxation for non-spherical samples
\cite{Ohm98,Prokofev98,Ohmthesis,Prokofev98JLTP}.

In Fig.~\ref{fig4}(b) we see the extraordinary effect of sample
annealing (i.e. for small values of the initial magnetization $M_{\rm
in}$) on the evolution of $P(\xi_H,H_t,t_t)$ with time. Now the
depletion proceeds over an extremely narrow bias range. This is
virtually incontestable experimental proof that we are seeing
tunneling relaxation. The narrowing of the hole is because in the
annealed sample further incremental relaxation hardly changes the
internal demagnetization field. Notice that the initial line shape of
$P(\xi_H)$ is very accurately fit to a Gaussian for the annealed
samples, exactly as predicted for the dipole field distribution of a
dense set of randomly oriented spins \cite{Berkov96}.

Further investigation of the effect of sample annealing led us to
another remarkable discovery (Fig.~\ref{fig5}). Progressive annealing
such that $\vert M_{\rm in}\vert<\vert 0.5M_s\vert$, eventually leads
to a hole linewidth which at short times is independent of further
annealing, and has a half linewidth of 0.8~mT. It is interesting that
such an intrinsic linewidth was predicted by Prokof'ev-Stamp
\cite{Prokofev98}. It is claimed to come from the nuclear spins which
would give rise to a linewidth $\xi_0$ of roughly the same order
(although only 2\% of natural iron has a nuclear moment, there are
other nuclei in the clusters that can contribute to the hyperfine
fields, i.e. more than 100 hydrogen, 18 nitrogen and 8 bromine atoms!). We
notice that any intrinsic linewidth due to the tunneling matrix
element itself is 5 orders of magnitude smaller, and would be quite
unobservable. According to Eq.~(\ref{eq2}), the ratio $\Gamma_{\rm
sqrt}/E_D$ (where $E_D$ is the Gaussian half-width of $P(\xi_H)$ for
strongly annealed samples) should be a constant, and thus allows us
to estimate $\Delta_{\rm tunnel}$ from our relaxation measurements. We
find that it is indeed a constant (even though $E_D$ and $\Gamma_{\rm
sqrt}$ vary with $M_{\rm in}$), and we extract $\Delta_{\rm
tunnel}\approx 5\times 10^{-8}$~K for $\vert M_{\rm in}\vert<\vert
0.5M_s\vert$, assuming $c=1$. This agrees well with the expected
value \cite{Ohm98,Prokofev98}.

In conclusion, we have developed a new measurement technique yielding
$P(\xi_H)$ which is related to the internal dipole field
distributions always present in crystals of molecular clusters. The
distribution evolves during relaxation by tunneling in a non-trivial
way, and can be monitored by our technique, revealing the details of
how the tunneling is proceeding in the sample, which molecules are
tunneling, and how the time-varying internal fields influence the
relaxation. The shape of the hole for thermal annealed distributions
indicates a fast dynamic relaxation over a field range of 0.8~mT
which could correspond in the Prokof'ev-Stamp theory to the nuclear
linewidth $\xi_0$. Although this is only indirect evidence of the
nuclear mechanism, it is hard for us to see what else could be
operating at these temperatures. Our evidence for the role of the
dipole interactions is on the other hand very direct, and in good
agreement with Monte Carlo simulations. We believe that our technique
should work for other multi-particle spin systems in the quantum
regime (like quantum spin glasses \cite{Thill95}), and could give
quite new information on the non-ergodic relaxation behavior typical
of these systems.

\begin{figure}
%\centerline{\epsfxsize=13 cm \epsfbox{fig1.eps}}
\caption{Magnetic hysteresis curves for a crystal of Fe$_8$ molecular
clusters in the quantum regime (40~mK) for three field sweeping
rates. Resonant tunneling is evidenced by six equally separated steps.
Note the strong dependence on the field sweeping rate. The
measurements were made using an array of micro-SQUIDs (see inset)
\protect\cite{Wernsdorfer97,Wernsdorfer96}. The high sensitivity of
this magnetometer allows us to study single crystals of the order of
10 to 500~$\mu$m which are placed directly on the array.}
\label{fig1}
\end{figure}

\begin{figure}
%\centerline{\epsfxsize=13 cm \epsfbox{fig2.eps}}
\caption{Typical square root of time relaxation curves for an Fe$_8$
crystal measured at 40~mK. For each curve, the sample was first
thermally annealed at $H$~= 0 (ZFC), then a small field was applied
and the relaxation of magnetization was measured during 1000~s. The
slope of the lines gives $\Gamma_{\rm sqrt}$ when plotted against the
square-root of $t$ as shown.}
\label{fig2}
\end{figure}

\begin{figure}
%\centerline{\epsfxsize=13 cm \epsfbox{fig3.eps}}
\caption{Field dependence of the short time square root relaxation
rates $\Gamma_{\rm sqrt}(\xi_H)$ for three different values of the
initial magnetization $M_{\rm in}$. According to Eq.~(\ref{eq2}), the
curves are proportional to the distribution $P(\xi_H)$ of magnetic
energy bias $\xi_H$ due to local dipole field distributions in the
sample. Note the logarithmic scale for $\Gamma_{\rm sqrt}$. The
peaked distribution labeled $M_{\rm in}=-0.998M_s$ was obtained by
saturating the sample, whereas the other distributions were obtained
by thermal annealing (FC or ZFC). $M_{\rm in}=0.870M_s$ is distorted
by nearest neighbor lattice effects.}
\label{fig3}
\end{figure}

\begin{figure}
%\centerline{\epsfxsize=13 cm \epsfbox{fig4.eps}}
\caption{Tunneling distributions: the field dependence of the short
time square root relaxation rates $\Gamma_{\rm sqrt}(\xi_H)$ are
presented on a logarithmic scale showing the depletion of the
molecular spin states by quantum tunneling at $H$~= 0 for various
tunneling times.\\
(a) Tunneling distributions for the initial magnetization starting
from saturation $M_{\rm in}=M_s$. For each point, the sample was first
saturated in a field of -1.4~T and at a temperature of about 2~K and
then rapidly cooled to 40~mK. After applying a ``tunneling field'' of
$H_t$~= 0, we let the sample relax for a ``tunneling time'' $t_t$
(after $t_t$~= 2, 10, 50 and 250~s, the reversed fraction of
magnetization is 0.008, 0.030, 0.077, 0.159 of $M_s$, respectively).
Finally, we applied a small field to measure the short time relaxation
which could be fit accurately to a square root law yielding
$\Gamma_{\rm sqrt}$. Because $\Gamma_{\rm sqrt}$ is proportional to
the spins which are still free to tunnel, one obtains the
distribution $P(\xi_H,H_t,t_t)$.
 (b) Tunneling distributions as in (a), but now for each point, the
sample was first annealed (FC) to a value $M_{\rm in}=-0.2M_s$. After
$t_t$~= 5, 10, 20 and 40~s, the reversed fraction of the magnetization is
0.0012, 0.0020, 0.0032, 0.0049 of $M_s$, respectively. At the
resonance, the depletion develops very rapidly with elapsed time,
even though the total magnetization and the internal demagnetization
field hardly change during this time. Notice that parabolic shape of
$\Gamma_{\rm sqrt}(\xi_H)$ shows it is accurately Gaussian, with a
half-width of 0.030~T. This Gaussian profile is found for $\vert
M_{\rm in}\vert<\vert 0.5M_s\vert$ (but with a half-width $E_D$
depending on $M_{\rm in}$).}
\label{fig4}
\end{figure}

\begin{figure}
%\centerline{\epsfxsize=13 cm \epsfbox{fig5.eps}}
\caption{Detail showing the depletion of molecular spin sates by
quantum tunneling for different values of annealing (FC). The
difference between the initial thermally quenched distribution (i.e.
$t_t$~= 0) and the tunneling distribution obtained after allowing the
sample to relax for $t_t$~= 16~s are plotted. For initial FC
magnetization values close to saturation, the depletion is larger and
asymmetric whereas it is narrow and symmetric for initial
magnetization values which are smaller than $\vert 0.5M_s\vert$. Note
that in the latter case the hole is independent of initial
magnetization, yielding an intrinsic hole width of about 0.8~mT (see
also fig.~\protect\ref{fig4}b).}
\label{fig5}
\end{figure}

\widetext

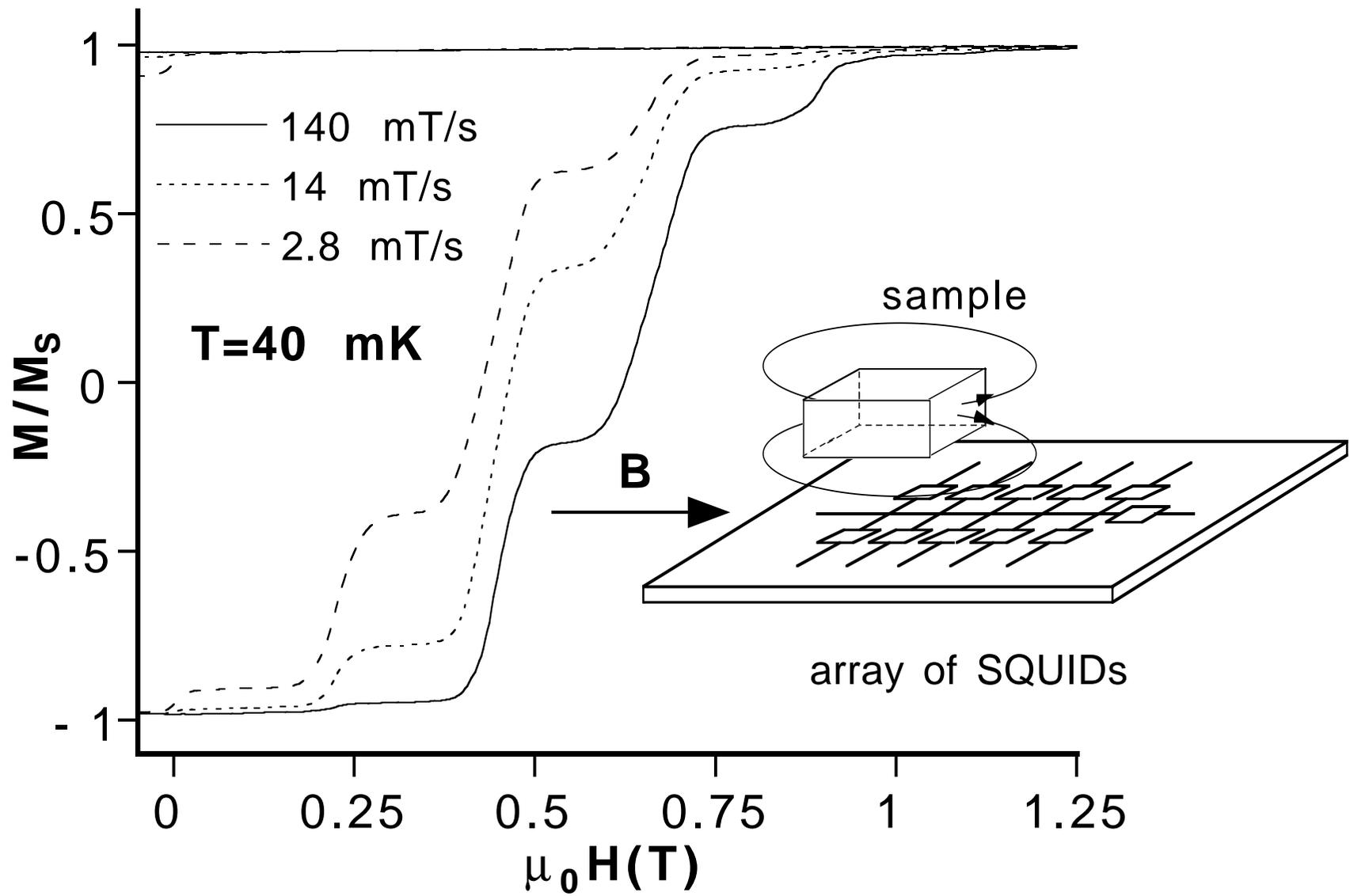
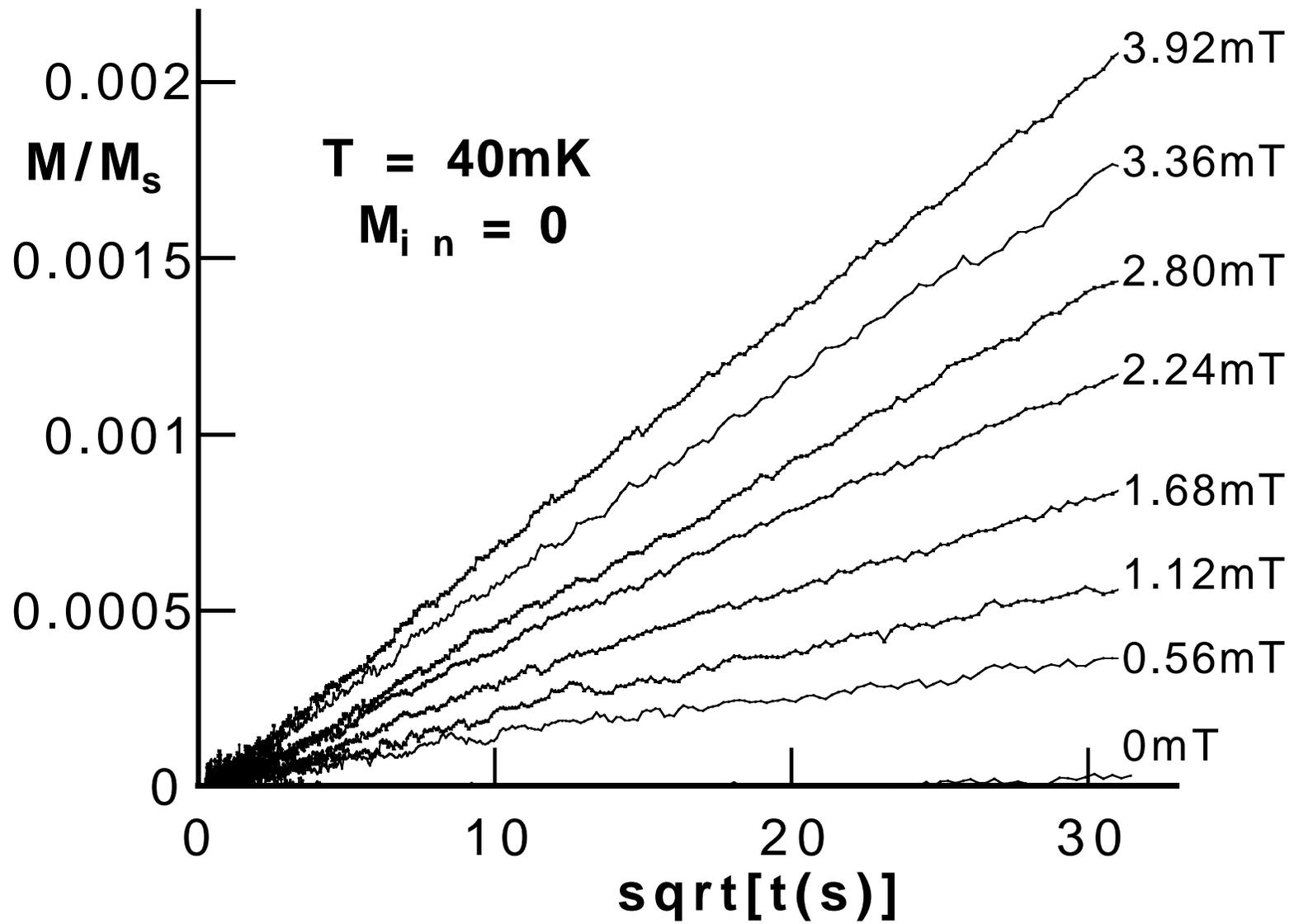
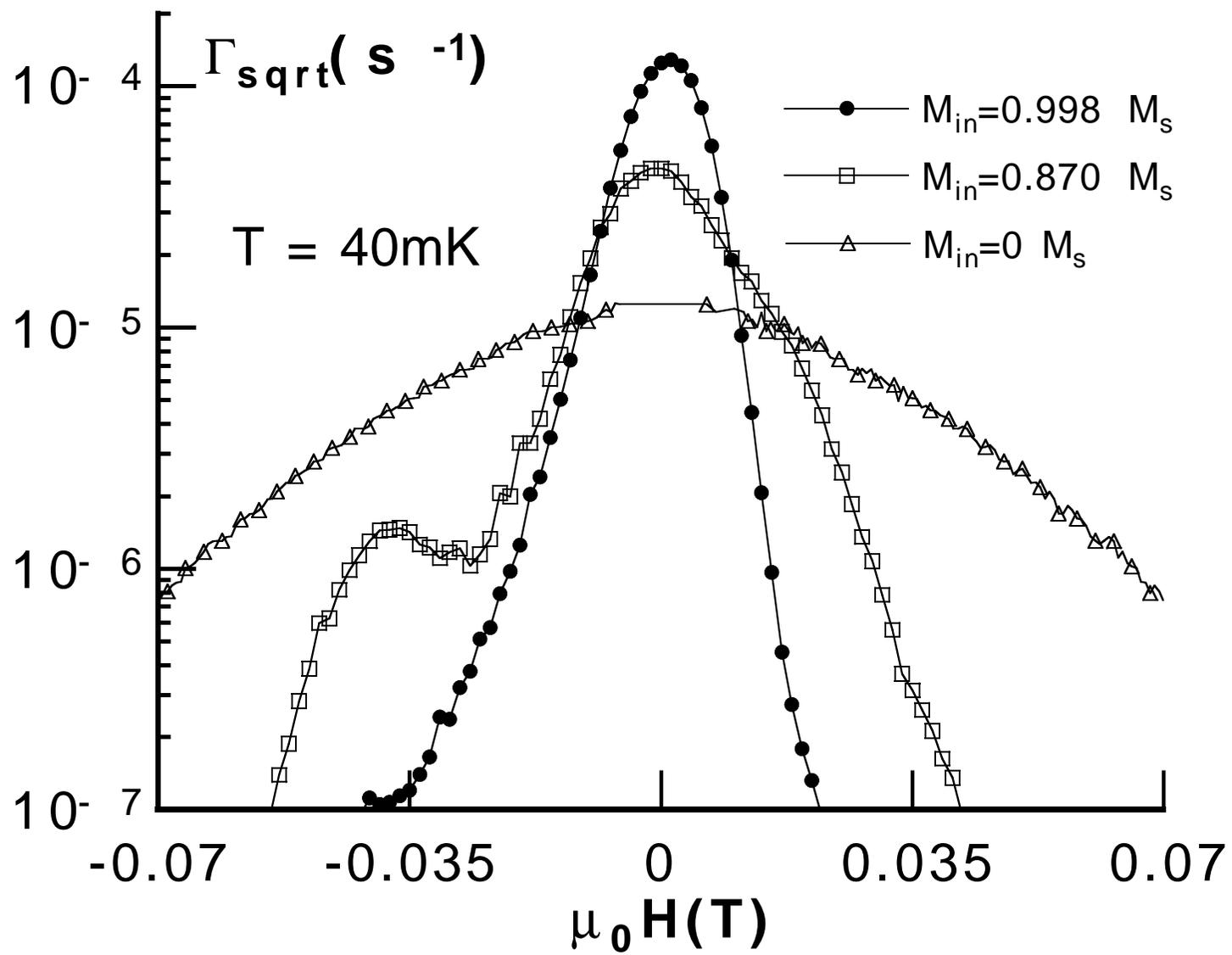
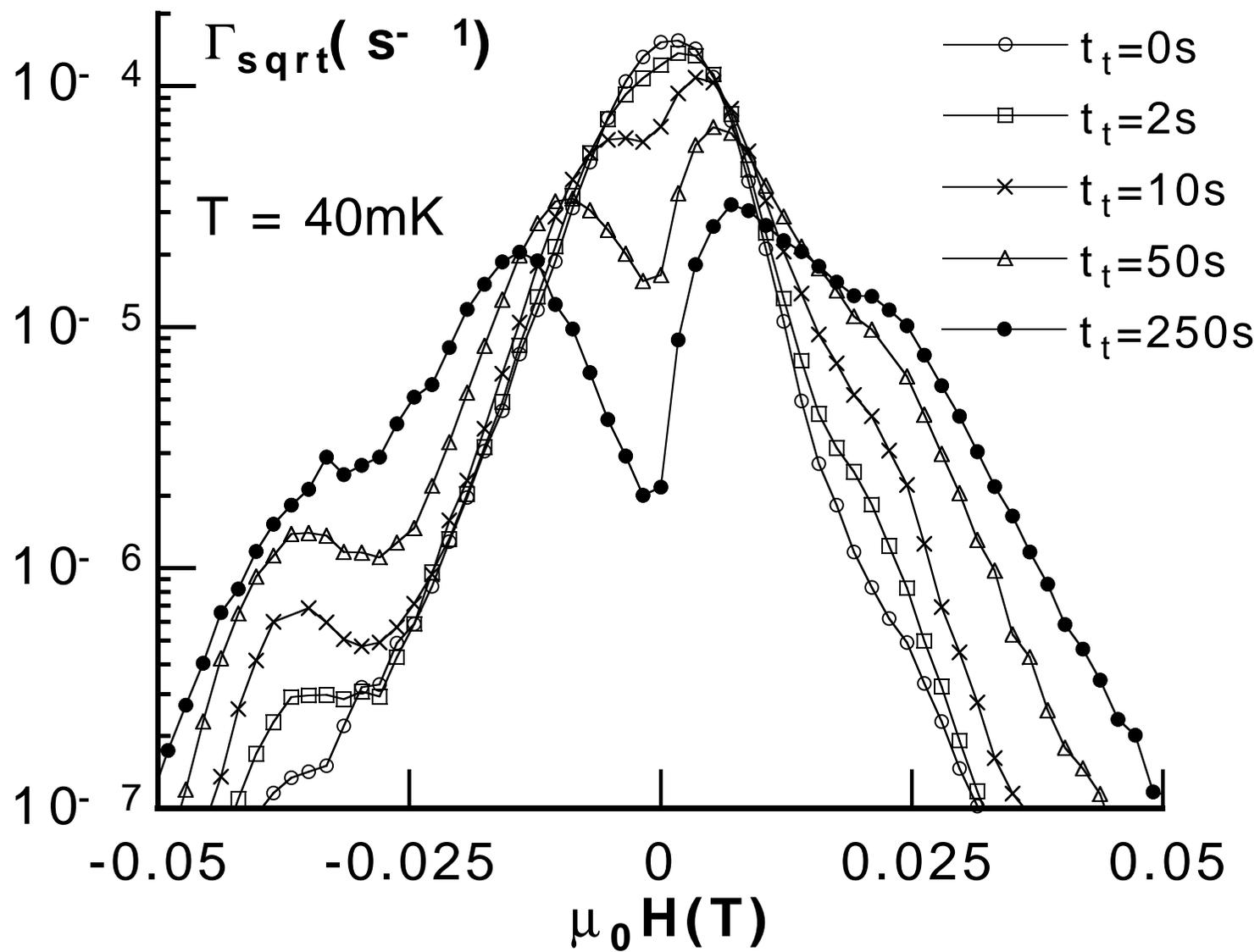
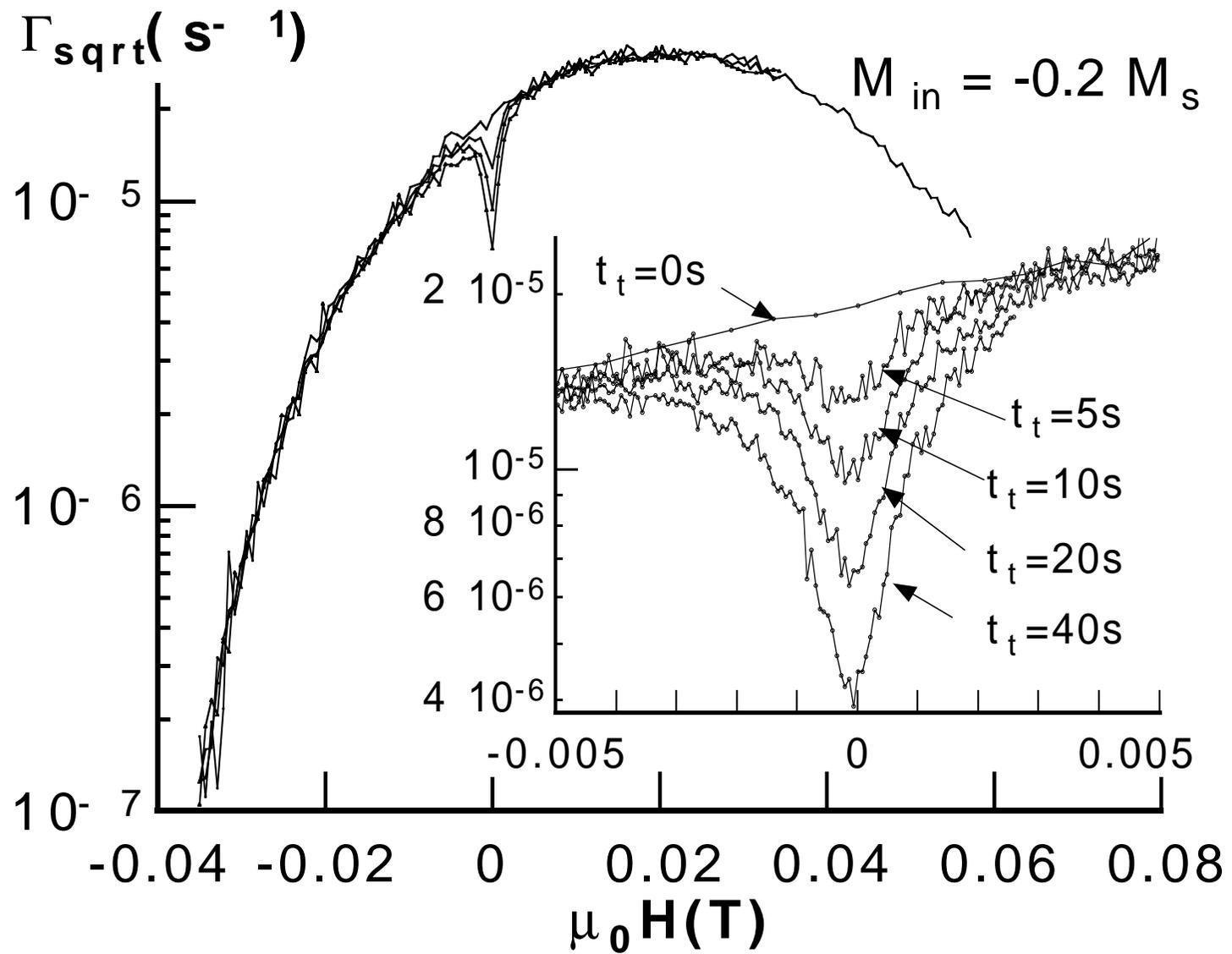
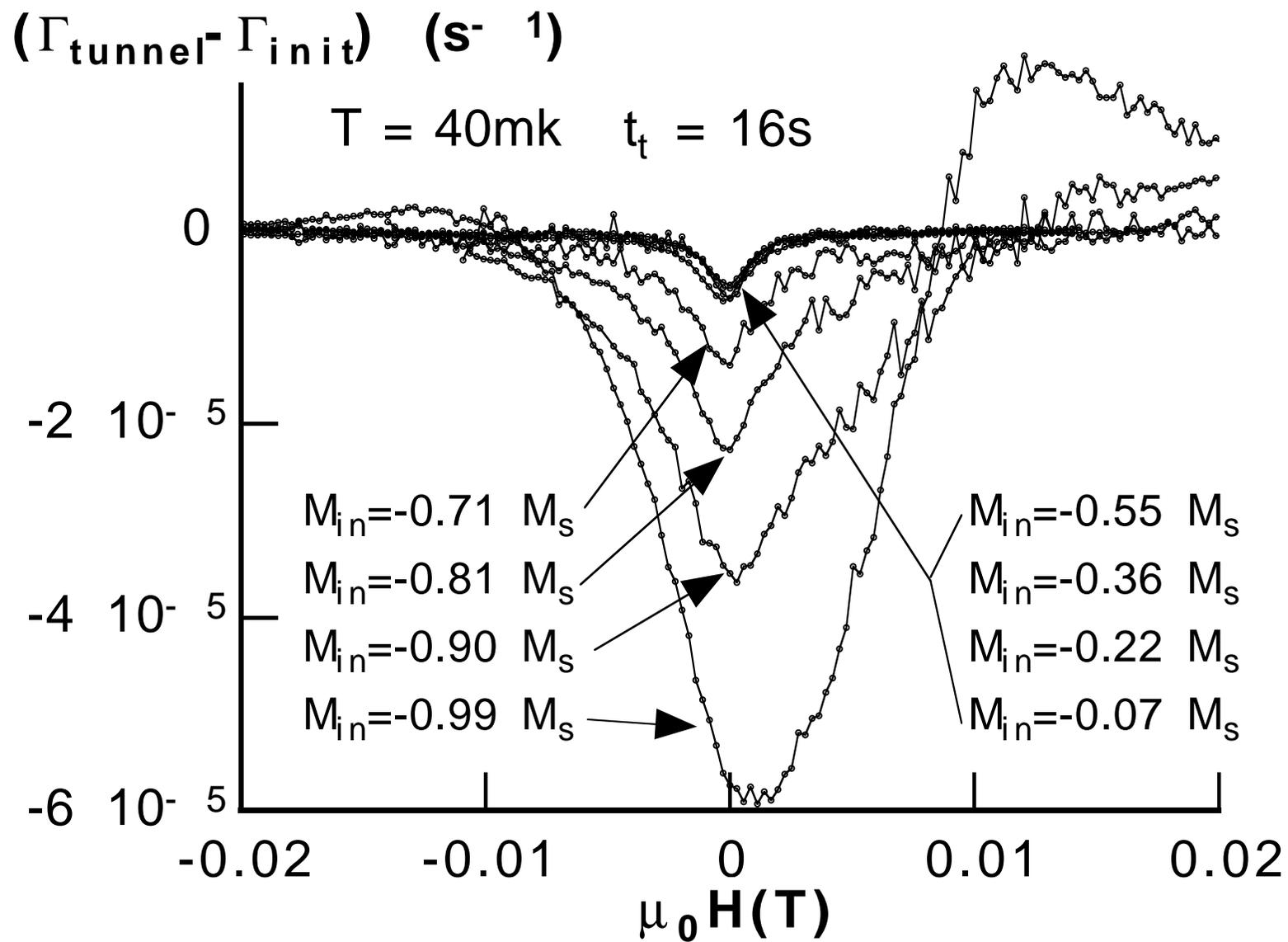
\end{document}